\documentclass{article}
\usepackage{spconf,amsmath,graphicx,hyperref}
\usepackage[utf8]{inputenc} 
\usepackage[T1]{fontenc}    
\usepackage{hyperref}       
\usepackage{url}            
\usepackage{booktabs}       
\usepackage{amsfonts}       
\usepackage{multirow}        
\usepackage{graphicx}       
\usepackage{nicefrac}       

\newcommand{\speech}{\texttt{[Speech]}}
\newcommand{\textmod}{\texttt{[Text]}}

\title{Efficient Interleaved Speech Modeling through Knowledge Distillation}
\name{Mohammadmahdi Nouriborji$^{\star}$ \qquad Morteza Rohanian$^{\dagger}$}
\address{
    $^{\star}$Nlpie Research\\
    $^{\dagger}$University of Zurich
}
\begin{document}
%
\maketitle
\begin{abstract}
Current speech language models exceed the size and latency constraints of many deployment environments. We build compact, expressive speech generation models through layer-aligned distillation, matching hidden states, attention maps, and softened logits to compress large multimodal transformers by 3× with minimal loss in performance. We introduce TinyWave, a family of 2B-parameter models for speech-to-speech and interleaved speech–text generation, trained on 50k hours of public audio. TinyWave supports (i) speech-only generation using phonetic or expressive tokens and (ii) mixed speech–text continuations. Evaluation on Libri-Light shows TinyWave within 1.4 normalized-perplexity points of its teacher.  Accuracy on spoken StoryCloze and SALMon reaches 93–97\% of the teacher’s performance, outperforming size-matched baselines. These models are optimized for deployment on commodity hardware, enabling applications in real-time conversational agents, assistive technologies, and low-resource environments. We release models, training code, and evaluation scripts to facilitate reproducible research on compact, expressive speech generation.\footnote{Generation samples are available at: \href{https://mohammadmahdinoori.github.io/tinywave-landing/}{https://mohammadmahdinoori.github.io/tinywave-landing/}; models are available on Hugging Face: \href{https://huggingface.co/tinywave}{https://huggingface.co/tinywave}; training and inference code is available at: \href{https://github.com/mohammadmahdinoori/TinyWave}{https://github.com/mohammadmahdinoori/TinyWave}.}
\end{abstract}
\begin{keywords}
speech generation, speech language models, knowledge distillation, multimodal transformers
\end{keywords}
\section{Introduction}
\label{sec:intro}

Expressive speech models aim to generate speech that captures not only the correct words, but also rich prosody, speaker-specific traits, and emotionally appropriate delivery. These features are essential for natural and effective human–computer interaction, particularly in storytelling, conversation, and assistive settings. While large language models (LLMs) excel at reasoning and linguistic generalization, text-only systems overlook acoustic and paralinguistic information critical to spoken communication. Speech models often struggle with semantic understanding due to the complexity of mapping discrete audio tokens to high-level linguistic concepts, a challenge compounded by limited paired speech-text data.

Recent research has explored multimodal models that process both speech and text, as well as interleaved inputs where audio and text alternate within a single context. These models better reflect real-world communication patterns and offer more flexible interaction. However, they typically require extensive training data and compute, making them unsuitable for edge deployment and resource-constrained environments.

A more tractable but limited alternative assembles an automatic speech recognition (ASR) frontend, a text-only LLM, and a text-to-speech (TTS) backend. While effective for basic language tasks, this pipeline discards much of the expressive signal leading to unnatural outputs. Recent speech models address this by modeling audio end-to-end using discrete units. Most of these models are trained from scratch at large scales, and model providers often maintain separate checkpoints for each deployment size. This practice is resource-intensive and inefficient, especially when targeting multiple operating points.

\begin{figure}[t]
    \centering
    \includegraphics[width=0.45\textwidth]{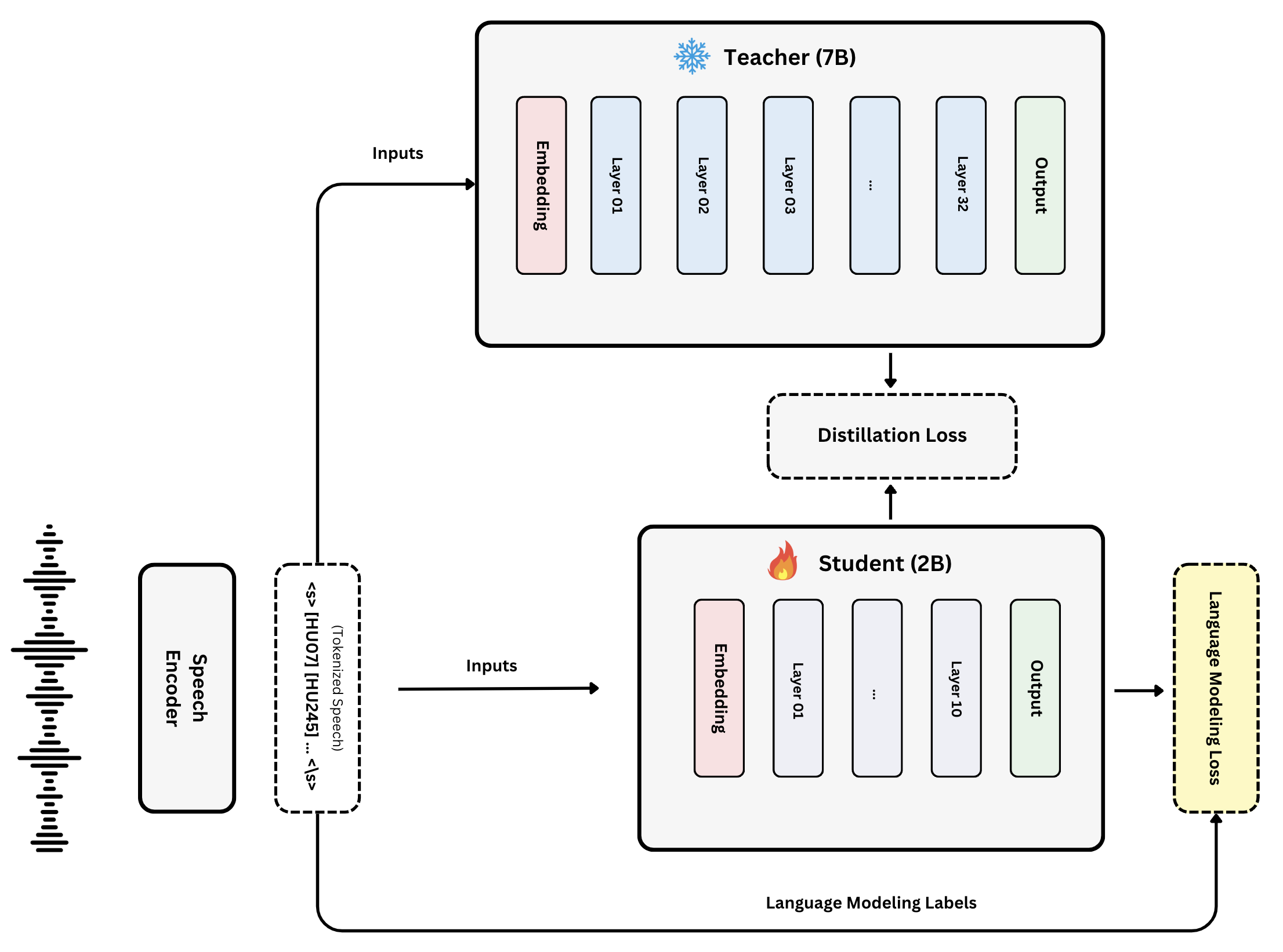}
    \caption{Distillation framework: The student aligns with the teacher’s logits, intermediate states, and ground-truth labels, guided by a multi-part loss.}
    \label{fig:distillation}
\end{figure}

We introduce TinyWave, a family of 2 billion-parameter speech-to-speech transformers distilled from the larger SpiritLM \cite{nguyen2025spiritlm}. TinyWave models preserve expressive qualities of speech and support both audio-only and interleaved speech–text inputs. Instead of training new models from scratch, we use layer-aligned distillation to transfer knowledge from a high-capacity teacher. We match hidden states, attention maps, and softened logits between teacher and student, encouraging the student to learn the teacher’s internal mechanisms. We also fine-tune the teacher on target-domain data to reduce distribution mismatch before distillation.

TinyWave performs strongly across several benchmarks. On StoryCloze and the SALMon style suite, it retains 93–97\% of the teacher’s accuracy. On Libri-Light continuations, it remains within 1.4 Normalised Perplexity Score (NPS) points and sometimes surpasses the teacher on style metrics. It also outperforms several larger models from prior work, showing that well-designed, compact models can rival or exceed larger systems under constrained resources.

All experiments use the publicly available Libri-Light dataset, which represents only a fraction of the data used to train the teacher model \cite{nguyen2025spiritlm}. Our results show that high-quality expressive speech modeling is possible without access to private datasets or large-scale compute.

\noindent In summary, our work makes the following contributions:
\begin{enumerate}
    \item A layer-aligned distillation framework for expressive end-to-end speech generation, enhanced by teacher fine-tuning to reduce domain shift.
    \item Three efficient models: \textsc{Base-Speech}, \textsc{Expressive-Speech}, and \textsc{Expressive-Interleaved}, covering audio-only and multimodal speech–text inputs.
    \item Open-source training and evaluation tools that enable reproducibility and deployment on commodity hardware.
\end{enumerate}

\section{Methodology}
\label{sec:method}

We present the TinyWave family: compact 2-billion-parameter speech-to-speech models distilled from 7B-parameter SPIRIT LM teachers \cite{nguyen2025spiritlm}. The SPIRIT LM models extend LLaMA-2 with speech capabilities by training on continuous streams of HuBERT-based audio tokens and interleaved text. TinyWave models inherit this cross-modal foundation through distillation. We release two speech-only variants and one interleaved model that handles mixed audio-text sequences. 

\subsection{Model Initialization} 
We initialize the TinyWave models to retain the core architectural properties of their SPIRIT LM teachers. For the speech-only models, \textsc{Base-Speech} and \textsc{Expressive-Speech}, we start from SpiritLM-Base-7B and SpiritLM-Expressive-7B, respectively. We prune every third transformer block, reducing the depth to 10 layers while keeping the original embedding layers, rotary position encoding (RoPE) settings, and language modeling head. This pruning yields models with approximately 2 billion parameters.

SPIRIT LM uses HuBERT-based discrete audio tokens, grouped into two variants: (i) a base phonetic tokenizer with 100 clusters and (ii) an expressive tokenizer that adds pitch and style tokens on top of the phonetic stream. The base speech variant adopts the base tokenizer, while expressive speech uses the expressive tokenizer to support richer acoustic modeling, including prosody and speaker variation.

For the interleaved model, \textsc{Expressive-Interleaved}, we initialize from the already distilled expressive-speech student rather than the 7B teacher. This transfer speeds up training and leverages the expressive speech priors learned during earlier distillation.

Teacher correction involves fine-tuning SpiritLM on a 10,000-hour subset of Libri-Light using LoRA adapters. This step aligns the teacher’s distribution with the target domain, reducing mismatch during distillation and improving supervision quality.

\subsection{Speech-Only Models} 
For the base speech variant, we fine-tune the SpiritLM-Base-7B teacher on a 20\% subset of Libri-Light. This teacher correction step ensures the supervision matches the target domain and avoids distributional mismatch. We skip this step for the expressive-speech model, since its teacher is already trained to encode prosodic and expressive cues relevant to the target.

We tokenize audio into sequences of HuBERT-based discrete units using the appropriate tokenizer. We train the student with an autoregressive language modeling loss, predicting the next token in the sequence. We apply our distillation framework to align the student’s internal representations and output distributions with the teacher’s, as described below.

\subsection{The Interleaved Model}
The expressive interleaved model extends speech modeling to mixed audio-text inputs. Following SPIRIT LM, we create interleaved training data by combining Libri-Light speech with pseudo-transcriptions generated by Whisper-v3 Large. We construct interleaved input sequences by inserting text spans into audio token streams at word boundaries. This preserves semantic continuity and minimizes alignment errors between modalities.

We sample from five interleaving patterns: \speech, \textmod, \speech\textmod, \textmod\speech, and \speech\textmod\speech, each with a fixed probability. This sampling encourages diversity and teaches the model to switch between modalities mid-sequence. We fine-tune the 7B expressive teacher on this interleaved corpus for a short duration (10k steps), then distill its knowledge into the student using our distillation method.

\subsection{Layer-to-Layer Distillation} 
\begin{table}[t!]
\centering
\label{tab:ppl_scores}
\small
\footnotesize 
\resizebox{\columnwidth}{!}{%
\begin{tabular}{llccc}
\toprule
\textbf{Type} & \textbf{Variant} & \textbf{GPT-2} & \textbf{LLaMA-1B} & \textbf{LLaMA-3B} \\
\midrule
\multirow{3}{*}{Base}
  & Teacher   & 0.93 & 0.95 & 0.95 \\
  & Distilled & 0.87 & 0.92 & 0.91 \\
  & Baseline  & 0.73 & 0.80 & 0.80 \\
\midrule
\multirow{3}{*}{Expressive}
  & Teacher   & 0.87 & 0.89 & 0.89 \\
  & Distilled & 0.85 & 0.88 & 0.87 \\
  & Baseline  & 0.75 & 0.78 & 0.77 \\
\bottomrule
\end{tabular}
}
\caption{Normalised Perplexity Score (↑).}
\end{table}

Our layer-to-layer distillation framework, shown in Figure~\ref{fig:distillation}, allows TinyWave models to learn from the internal structure of their larger teachers. We supervise student models at three levels: internal representations, output predictions, and autoregressive generation.

We align hidden states and attention maps by mapping each student layer \( l \) to a teacher layer using \( g(l) = 3l + 4 \). The alignment loss is defined as:

\begin{equation}
\mathcal{L}_{\text{align}} = \sum_{l} \alpha_l \mathcal{L}_{\text{cos}}(h_{g(l)}^{(t)}, h_l^{(s)}) + \gamma_l \text{KL}(A_{g(l)}^{(t)} \| A_l^{(s)})
\end{equation}

Here, \( h_{g(l)}^{(t)} \) and \( h_l^{(s)} \) are hidden states from the teacher and student, respectively. \( \mathcal{L}_{\text{cos}} \) enforces directional similarity, while KL divergence between attention maps ensures structural alignment. Scalars \( \alpha_l \) and \( \gamma_l \) weight each term appropriately.

We align output distributions using softened logits with temperature \( \tau \):

\begin{equation}
\mathcal{L}_{\text{output}} = \text{KL}\left( \text{softmax}\left(\frac{y^{(t)}}{\tau}\right) \hspace{2.5pt} \| \hspace{5pt} \text{softmax}\left(\frac{y^{(s)}}{\tau}\right) \right)
\end{equation}

This encourages the student to approximate the teacher’s full output distribution, not just the argmax predictions.

Finally, we apply a cross-entropy loss over ground-truth tokens:

\begin{equation}
\mathcal{L}_{\text{LM}} = - \sum_t \log P(y_t \mid y_{<t})
\end{equation}

The full objective combines all three components:

\begin{equation}
\mathcal{L}_{\text{total}} = \lambda_1 \mathcal{L}_{\text{align}} + \lambda_2 \mathcal{L}_{\text{output}} + \lambda_3 \mathcal{L}_{\text{LM}}
\end{equation}

This loss enables the student to replicate the teacher’s behavior efficiently, making TinyWave suitable for deployment in constrained environments without compromising expressive generation capabilities.

\begin{table}[t!]
\centering
\label{tab:ling_eval}
\small
\footnotesize 
\setlength{\tabcolsep}{5pt}
\begin{tabular}{lcccc}
\toprule
\textbf{Model} & \textbf{sStory} & \textbf{tStory} & \textbf{sWUGGY} & \textbf{sBLIMP} \\
\midrule
GSLM                     & 53.3 & 66.6 & 64.8 & 54.2 \\
AudioLM                  &  –   &  –   & 71.5 & \textbf{64.7} \\
VoxtLM                   &  –   &  –   & 66.1 & 57.1 \\
TWIST                    & 55.4 & 76.4 & \textbf{74.5} & 59.2 \\
\midrule
Teacher (7 B)            & \textbf{60.3} & \textbf{82.1} & 70.98 & 58.37 \\
TinyWave-Base (2 B)      & 55.3 & 74.8 & 70.26 & 58.08 \\
Baseline (2 B)           & 53.6 & 53.6 &  68.72   &  52.85   \\
\midrule
Teacher-Expressive       & 57.1 & 75.5 & 65.00 & 54.35 \\
TinyWave-Expressive      & 54.1 & 71.6 & 63.98 & 54.28 \\
Baseline-Expressive      & 50.6 & 63.1 & 67.35 & 52.42 \\
\bottomrule
\end{tabular}
\caption{Linguistic reasoning in speech.  
Higher is better. Bold marks the best score in each column.}
\end{table}

\section{Results}
\label{sec:results}


We evaluate TinyWave across multiple dimensions: (i) language modeling quality (NPS), (ii) spoken commonsense reasoning (StoryCloze), (iii) morphosyntactic competence (sWUGGY, sBLIMP), and (iv) acoustic consistency and alignment (SALMon) \cite{maimon2025salmon}. Throughout, Teacher refers to the original 7 B SpiritLM checkpoint, Distilled to the 2 B TinyWave student, and Baseline to an equally sized model trained from scratch without distillation. In addition to the teacher model, we compare our approach with several strong baselines including GSLM \cite{lakhotia2021librilight}, AudioLM \cite{borsos2023audiolm}, VoxtLM \cite{maiti2023voxtlm}, TWIST \cite{hassid2023twist}, and LAST-1.3 \cite{turetzky2025last13}.

\begin{table*}[t]
\centering
\label{tab:salmon_results}
\small  
\begin{tabular}{lccccc}
\toprule
\textbf{Metric} & \textbf{Teacher} & \textbf{TinyWave-Expr.} & \textbf{LAST-1.3B} & \textbf{TWIST-1.3B} & \textbf{TWIST-7B} \\
\midrule
Background Domain   & 55.0 & \textbf{57.0} & 56.0 & 55.5 & 55.0  \\
Background All      & 64.0 & 58.5          & 61.0 & 60.5 & 60.5  \\
Gender Consistency  & 85.0 & 76.5          & 68.5 & 69.5 & 70.0  \\
Speaker Consistency & 81.0 & 71.0          & 64.5 & 69.0 & 71.0  \\
Sentiment Alignment & 52.0 & 51.0          & 53.5 & 53.0 & 51.5  \\
\bottomrule
\end{tabular}
\caption{SALMon accuracy (↑).}
\end{table*}

\subsection{Language-Modeling Quality} 
Table \ref{tab:ppl_scores} lists Normalised Perplexity Scores (NPS; higher is better). Distillation narrows the gap to the teacher by 85--93 \% while halving parameter count. The expressive variants score 2--3 pp lower than their base counterparts, reflecting the added burden of prosody tokens. Baselines trail both distilled and teacher models by a wide margin, confirming that the student networks benefit from teacher guidance rather than sheer capacity.

\subsection{Spoken Commonsense Reasoning}
Spoken StoryCloze evaluates whether a model chooses the coherent ending for a four-sentence prompt. TinyWave-Base achieves 55.3 \% (sStoryCloze) and 74.8 \% (topic variant), improving over the 53.6 \% baseline and retaining most of the teacher’s advantage (Table \ref{tab:ling_eval}). Expressive-speech trails by roughly two points, consistent with its slightly higher linguistic entropy.

\subsection{Surface-Level Morphosyntax}
Table~\ref{tab:ling_eval} presents accuracy on sWUGGY and sBLIMP, which evaluate surface-level linguistic competence in generated speech. sWUGGY tests the model’s lexical sensitivity by measuring its ability to distinguish plausible from implausible wordforms. sBLIMP probes grammatical knowledge through syntactic minimal pairs. TinyWave-Base closely tracks the teacher, scoring 70.26\% on sWUGGY and 58.08\% on sBLIMP. TinyWave-Expressive yields slightly lower scores, consistent with its expressive modeling focus. Both models outperform GSLM and VoxtLM, showing that distillation preserves morphosyntactic structure despite aggressive model compression.

\subsection{Acoustic Consistency and Alignment}
SALMon probes fine-grained acoustic factors. TinyWave-Expressive keeps most of the teacher’s scores, matching or exceeding it on background-domain tests, while dropping 8--10 pp on speaker-specific metrics (Table \ref{tab:salmon_results}). TinyWave’s superior performance on background-domain tests may stem from distillation’s regularization effect, which reduces overfitting to speaker-specific features and enhances generalization to diverse acoustic contexts. Distillation therefore preserves global acoustic cues better than speaker identity; we expose this trade-off for future optimisation.

\section{Related Works}

LLMs have recently been extended to operate on waveform tokens. This enables high-quality multi-speaker speech synthesis \cite{wang2023valle, kharitonov2023multispk, jiang2024multispk, ji2024speech} and general audio generation \cite{kreuk2023audiogen, borsos2023audiolm}. Models like VALL-E and AudioLM separate phonetic and acoustic information to improve fidelity \cite{wang2023valle, borsos2023audiolm}.

Self-supervised speech models—such as wav2vec 2.0, HuBERT, and w2v-BERT—can be discretized into pseudo-phonemes. These discrete tokens allow LMs to learn over raw audio \cite{lakhotia2021librilight}. The tokens capture both linguistic content and prosody \cite{kharitonov2022prosody}, supporting tasks like emotion conversion \cite{kreuk2023audiogen} and spoken dialogue generation \cite{nguyen2023dialogue}. However, under equal data budgets, speech-only LMs still lag text LMs in semantic accuracy \cite{nguyen2023dialogue}.

GSLM \cite{lakhotia2021librilight} marked the start of large-scale generative speech models. Successors such as Spectron \cite{nachmani2023spectron}, AudioPaLM \cite{rubenstein2023audiopalm}, VioLA \cite{wang2023viola}, VoxtLM \cite{maiti2023voxtlm}, and SUTLM \cite{chou2023sutlm} unify ASR, TTS, and spoken continuation. Despite impressive capability, these models exceed 7B parameters and are compute-heavy.

Knowledge distillation addresses this bottleneck. Originally, it involved matching softened output logits \cite{hinton2015distilling}. Later approaches aligned hidden states and attention maps \cite{sanh2019distilbert, minitron2024}. In some cases, student models outperform their teachers due to improved regularization. In speech, most distillation efforts have focused on ASR encoders \cite{huang2023distilwhisper}. TinyWave compresses large audio LLMs into a 2B student using a layer-wise distillation scheme. It retains near-teacher quality while enabling efficient inference on commodity hardware.

\section{Limitations}
\label{sec:limitations}



Despite the promising performance of TinyWave models across semantic and expressive benchmarks, several limitations remain. Notably, the models exhibit constrained semantic understanding, often prioritizing phonetic accuracy over deeper linguistic comprehension. This limitation is consistent with findings in self-supervised speech models, which encode phonetic information more robustly than semantic content. 

The current tokenizer and vocoder components present challenges in producing natural, human-like speech. Existing tokenizers may inadequately capture the stylistic nuances of conversation, leading to outputs that lack expressiveness and emotional depth. Vocoders often struggle with prosody resulting in synthetic speech that sounds flat or unnatural. Future work should explore the integration of advanced tokenization methods, such as Residual Vector Quantization (RVQ), and more sophisticated vocoders capable of modeling fine-grained prosodic features to improve the naturalness and expressiveness of generated speech.

\section{Conclusion}
\label{sec:conclusions}

TinyWave compresses a 7 B speech–text foundation model into 2 B parameters without discarding expressive power.  Layer-aligned distillation, anchored by a brief teacher correction phase, transfers both semantic reasoning and prosodic control to the student.  Across NPS, StoryCloze, SALMon, and ASR/TTS, TinyWave matches or closely trails its teacher and consistently outperforms size-matched baselines.  These results show that careful distillation, not additional scale, can deliver practical, high-quality speech generation on modest hardware.  We release all checkpoints and training code to support reproducibility and future work.

\vfill\pagebreak

\section{Acknowledgements}
No funding was received for conducting this study. The authors have no relevant financial or nonfinancial interests to disclose.

\section{Compliance with Ethical Standards}
This is a numerical simulation study for which no ethical approval was required.

\bibliographystyle{IEEEbib}

\end{document}